\documentclass[12pt]{article}
\usepackage{amssymb,latexsym}
\usepackage[cmex10]{amsmath}
\usepackage[all]{xy}
\input{xy}
\xyoption{all}
\title{On Rueppel's Linear Complexity Conjecture}
\author{Graham H. Norton
\footnote{
School of Mathematics and Physics, University of Queensland, Brisbane, Queensland 4072, Australia. (Email: ghn@maths.uq.edu.au)}\\\\
To Gladys Jackson (1922-2012) in loving memory}
\newtheorem{theorem}{\bf Theorem}[section]
\newtheorem{corollary}[theorem]{\bf Corollary}
\newtheorem{proposition}[theorem]{\bf Proposition} 
\newtheorem{notation}[theorem]{\sc Notation} 
\newtheorem{definition}[theorem]{\bf Definition} 
\newtheorem{lemma}[theorem]{\bf Lemma} 
\newtheorem{example}[theorem]{\it Example}  
\newtheorem{examples}[theorem]{\it Examples}  
\newtheorem{algorithm}[theorem]{\bf Algorithm} 
\newtheorem{conjecture}[theorem]{\bf  Conjecture}
\newtheorem{remark}[theorem]{\it Remark}
\newtheorem{remarks}[theorem]{\it  Remarks}

\newenvironment{proof}{{\noindent\it Proof.\ }}{$\square$\par\vspace{4mm}} 
\def \bt{ \begin{theorem} }
\def \et{ \end  {theorem} }
\def \bl{ \begin{lemma} }
\def \el{ \end  {lemma} }
\def \bp{ \begin{proposition} }
\def \ep{ \end  {proposition} }
\def \bn{ \begin{notation} }
\def \en{ \end  {notation} }
\def \bq{ \begin {question} }
\def \eq{ \end {question} }
\def \bc{ \begin{corollary} }
\def \ec{ \end  {corollary} }
\def \bcj{ \begin{conjecture} }
\def \ecj{ \end  {conjecture} }
\def \bd{ \begin{definition} }
\def \ed{ \end  {definition} }
\def \bdp{ \begin{definitionprop} }
\def \edp{ \end  {definitionprop} }
\def \bdt{ \begin{definitiontheorem} }
\def \edt{ \end  {definitiontheorem} }
\def \bpr{ \begin  {proof} }
\def \epr{ \end  {proof} }
\def \ba{ \begin{algorithm} }
\def \ea{ \end{algorithm} }
\def \be{ \begin{example} }
\def \eex{ \end{example} }
\def \bes{ \begin{examples} }
\def \eexs{ \end{examples} }
\def \br{ \begin{remark} }
\def \er{ \end{remark} }
\def \brs{ \begin{remarks} }
\def \ers{ \end{remarks} }
\def \bpb{ \begin{problem} }
\def \epb{ \end{problem} }

\newcommand{ \GF} {\mathrm{GF}}

\newcommand{\la } {\leftarrow}

\newcommand{\ol} {\overline}

\newcommand{\ul} {\underline}

\newcommand{\ra} {\rightarrow}

\newcommand{\vv} {\mathrm{v}}

\newcommand{\LT} {\mathrm{LT}}
\newcommand{\N} {\mathbb{N}}
\newcommand{\F} {\mathbb{F}}
\newcommand{\bbP} {\mathbb{P}}

\newcommand{\R} {\mathrm{R}}

\newcommand{\Q}{\mathbb{Q}}
\newcommand{\Z}{\mathbb{Z}}
\newcommand{ \K} {\mathbb{K}}

\newcommand{\I} {\mathcal{I}}

\newcommand{\LL} {\mathcal{L}}

\newcommand{\M} {\mathrm{M}}

\begin{document}
\maketitle
\begin{abstract}
Rueppel's conjecture on the linear complexity of the first $n$ terms of the sequence $(1,1,0,1,0^3,1,0^7,1,0^{15},\ldots)$  was first proved by Dai using the Euclidean algorithm. We have  previously shown that we can attach a homogeneous (annihilator) ideal of $\F[x,z]$ to the first $n$ terms of a sequence over $\F$ and construct a pair of generating forms for it. This approach gives another proof of Rueppel's conjecture.  We also prove additional properties of these forms  and deduce the  outputs of the LFSR synthesis algorithm applied to the first $n$ terms. Further,  dehomogenising the leading generators yields the minimal polynomials of Dai.

{\small {\bf Keywords:} Annihilator, finite sequence, form, homogeneous ideal,  Macaulay's inverse system, linear complexity.}
\end{abstract}
\maketitle
\section{Introduction}
\subsection{Background}
It was conjectured  in \cite{Rueppel} (and verified for $n\leq 128$ by direct calculation) that the binary sequence  $(1,1,0,1,0^3,1,0^7,1,0^{15},1,\ldots)$ has a perfect linear complexity profile i.e. for any $n$, the linear complexity of  the first $n$ terms  is $\lfloor (n+1)/2\rfloor$. According to  \cite[p.441]{Dai}, this was verified by Massey for $n=2^k-1$ and $n=2^k$. The conjecture  was first proved by Dai in \cite{Dai}. Her proof uses the Euclidean Algorithm (EA) and an element $\rho$ in  a quadratic  extension of $\GF(2)(x)$, which was not motivated. The proof in \cite{Dai} also requires an additional property of the EA, namely \cite[Proposition 2]{Dai}, proved in \cite[Lemma 5]{DW}. 

For another proof applying the continued fraction algorithm to the formal (inverse) power series of the sequence in  $\GF(2)[[x^{-1}]]$, see   \cite[Corollary 2]{Nied87}.  The partial quotients are always $x$. However in \cite{Nied86a}, \cite{Nied87} there are  convergence and irrationality considerations,  with denominators and numerators  obtained using polynomial division. 

The LFSR synthesis algorithm is from \cite{Ma69}; see also  \cite[Section 9.9]{DH}. A derivation of the 'connection polynomial'  for $n=2^k$ via the LFSR synthesis algorithm  appeared in \cite[pp. 46--47]{R86}. However,  \cite{R86} {\em assumes the essential result}: the $k^{th}$ 'discrepancy' $\Delta_k$ is  the residue class of $k$ in $\GF(2)$. As far as we know, the conjecture has not been completely proved using the LFSR synthesis algorithm (but see Corollary \ref{lfsr}).

When $2t$ terms of a sequence and an upper bound  $t$ on its linear complexity are known, the decoding algorithm of Berlekamp \cite[Algorithm 7.4]{Be68} can be used to find the linear complexity of a sequence. Likewise for the minimal polynomial algorithms of \cite[pp. 441--444]{LN} and  \cite{Fitz}. However, none of these  algorithms determines the linear complexity of a sequence {\em ab ovo}; see e.g. Example \ref{Fitzex}.
 We note that \cite[Algorithm 4.2]{MR} determines a sequence of rational approximations without assuming  an upper bound $t$ on the linear complexity and $2t$ terms of the sequence.
\subsection{Synopsis}
Let $n\geq1$ and $(r_0,\ldots,r_{n-1})$ be the first $n$ terms of the sequence.
We give a self-contained proof of the conjecture by working directly with $(r_0,\ldots,r_{n-1})$ and studying its 'annihilator ideal' $\I_{n-1}$ as defined in \cite{N15b}. The corresponding algorithm  is  easier to understand and remember than the LFSR synthesis algorithm of \cite{Ma69}. Our main result  is that in our inductive,  algorithm-free   construction, $\Delta_k=\ol{k}\in\GF(2)$ or equivalently, the 'intermediate shift' is $x$ if and only if  $k$ is odd, Theorem \ref{main}. This yields  generator forms of $\I_{n-1}$\,, a proof of the conjecture,  additional properties of these ideals and a short proof of the main theorem of \cite{Dai}, Corollary \ref{omnibus}. We use $(r_0,\ldots,r_9)$ and its inverse form $R^{(-9)}$ as  running example, see Example \ref{first8}.

Our approach is simpler: (i)  it is inductive, adapting to the next term of the sequence, so  we do not need to begin with all of $(r_0,\ldots,r_{n-1})$ as in \cite{Dai}  (ii) we do not assume the existence of an element $\rho$ as in \cite{Dai} (iii)  we work with forms in $\F[x^{-1},z^{-1}]$ and $\F[x,z]$, so there are no convergence or irrationality considerations  (iv) we work with denominators only (v) we do not assume an upper bound $t$ on the linear complexity of the sequence and $2t$ terms of the sequence. Our 'intermediate shifts' are analogous to the partial quotients of \cite{Nied86a}, \cite{Nied87}. The resulting algorithm for $(r_0,\ldots,r_{n-1})$   is division-- and multiplication--free and uses little more than a parity test on $n$; see Part (a) below and Algorithm \ref{Ralg}. 

In more detail, the preliminary Section \ref{prelim} concerns Macaulay's 'inverse system'  in two variables and how we adapt it.  Let $\F$ be any field. The  $\F[x,z]$ module $\F[[x^{-1},z^{-1}]]$ of 'inverse series' in variables $x^{-1},z^{-1}$ is known as Macaulay's inverse system (in two variables). Now  $\F[x^{-1},z^{-1}]$ is an $\F[x,z]$  submodule of $\F[[x^{-1},z^{-1}]]$; see Definition \ref{mod}, \cite{Northcott}. An 'inverse form' i.e. a form  $F\in\F[x^{-1},z^{-1}]^\times$ thus has a (homogeneous) annihilator ideal\footnote{The inverse form of $(s_0,\ldots,s_{n-1})$ was called its {\em generator form}  and its annihilator ideal was defined  differently in \cite{AD}; in \cite[Appendix]{N15b} we showed that the two ideals coincide.} $\I_F\subseteq\F[x,z]$. This  elementary algebraic structure underpins  our approach to the LFSR synthesis algorithm, with finite sequences replaced by inverse forms and characteristic polynomials replaced by annihilating forms, Theorem \ref{bij}.

Section \ref{VOP} recalls the inductive construction of \cite{N19}: i.e. the  construction of  a 'viable ordered pair (VOP)'  of forms  $(f,g)$ which generate $\I_F$\,;  here $z$ does not divide the graded lexicographic (grlex) leading term of $f$ (we call $f$ a 'leading minimal form'), $z$ divides $g$ and $|f|+|g|=2-|F|$, where $|\ |$ denotes total degree. This was further simplified and extended in \cite{N19}; see Theorem \ref{constr}, the corresponding $\mathcal{O}(|F|^2)$  algorithm and an example of it at work over $\Q$, Example \ref{Fitzex}.

We also include several new features of the construction e.g. the set of leading forms of $\I_F$ of minimal degree and the fact that $f$, $g$ are always relatively prime.
 
Section \ref{GF2} is devoted to $\F=\GF(2)$ and Rueppel's conjecture. We apply our construction to obtain a VOP  of  $\I_{n-1}$\,. 
  Our inductive basis  is $(f^{(0)},g^{(0)})=(x+z,z)$ and we prove that for $k\geq0$

(a)  the essential result $\Delta_k=\ol{k}\in\GF(2)$; equivalently, either we have (i) $(f^{(k+1)},g^{(k+1)})=(xf^{(k)}+g^{(k)},zf^{(k-1)})$ if $k+1$ is even or (ii) $(f^{(k+1)},g^{(k+1)})=(f^{(k)},zg^{(k)})$

(b)  $(f^{(k+1)},g^{(k+1)})$ is a VOP for $\I_{k+1}$
 
(c)   $|f^{(k+1)}|=\lfloor (k+3)/2\rfloor$  is the linear complexity of $(r_0,\ldots,r_{k+1})$

(d) either there is a unique leading form of minimal degree viz. $f^{(k+1)}$ (if $k+1$ is odd) or the leading forms of minimal degree are precisely $f^{(k+1)}$ and $f^{(k+1)}+g^{(k+1)}$ 

(e) $f^{(k+1)}(0,1)=g^{(k+1)}(0,1)=1$, so the reciprocal of $f^{(k+1)}(x,1)$ has degree $|f^{(k+1)}|$.

  Part (c) proves Rueppel's conjecture.  Parts (c), (e) imply that the  LFSR  Synthesis algorithm applied to $(r_0,\ldots,r_{2k-1})$ outputs  $k$ and the reciprocal of $f^{(2k-1)}(x,1)$.  Part (a) implies a recurrence for $f^{(2k-1)}(x,1)$ with characteristic polynomial $Y^2+xY+1\in\F(x)[Y]$, motivating the use of its roots $\rho^{\pm1}$ in an extension of $\F(x)$. Solving this recurrence yields
$$x\,f^{(2k-1)}(x,1)=(1+\rho)\rho^k+(1+\rho^{-1})\rho^{-k}.$$

 Finally, for $k\geq1$, let  $c_k\in\F[x]$  be found for $(r_0,\ldots,r_{2k-1})$ using the EA as in \cite{Dai}. Lemma 3, {\em op. cit.} implies that  $c_k$ satisfies this same recurrence, so that  $c_k(x)=f^{(2k-1)}(x,1)$.
 This also simplifies the proof of the main theorem of \cite{Dai} and implies that 
 if $k\geq0$ and $l=2^k$, then  $$f^{(2l-1)}(x,z)=x^l+\sum_{j=0}^{\log_2 l}x^{l-2^j}z^{2^j}.$$

 \section{Preliminaries}\label{prelim}
\subsection{Notation}
 In general, $\F$ is an arbitrary field and $\R=\F[x,z]$.  Multiplication in $\R$ is written as juxtaposition. For $\varphi,\varphi'\in\R$ and $k\in\N^\times$,  $x^k\,\varphi+\varphi'$ means $x^k\varphi(x,z)+\varphi'(x,z)$ and similarly for  $\varphi+\varphi'\,z^k$\,.  The total degree of $\varphi\in\R^\times$ is $|\varphi|$,  with $|x|=|z|=1$. 
The ideal  generated by $\varphi,\psi\in\R$ is written $\langle \varphi,\psi\rangle$.

For logical statements $P,Q,A,B$, we will often write $(P\|Q)$ if $(A\|B)$ as shorthand for ($P$ if $A$) and ($Q$ if $B$). For example, 'the intermediate shifts are $(x^d\|1)$ if $k$ is (odd$\|$even)'.

We also include reference tables of commonly-used symbols to aid the reader.

\begin{center}
\begin{tabular}{|c|l|}\hline
Symbol & Meaning \\\hline\hline
$\R$& $\F[x,z]$\\\hline
$d_k$ &$|g^{(k)}|-|f^{(k)}|\in\Z$\\\hline
$|f|$& the total degree of a form $f\in\R^\times$\\\hline
$(f^{(k)},g^{(k)})$ & the constructed viable  ordered pair (VOP) for $\I_k$\\\hline
$F$ & an inverse form in $\M^\times$\\\hline
$\F$& a field\\\hline
$\I_F$ & the annihilator ideal of an inverse form $F$\\\hline
$k$ & intermediate variable\\\hline
$\LL$ & the set of monic, non-zero (leading) forms $\varphi$ with $z\nmid\LT(\varphi)$.\\\hline
$\M$& the $\R$-module $\F[x^{-1},z^{-1}]$\\\hline
$n$ & length of $(s_0,\ldots,s_{n-1})$\\\hline
$(r_0,r_1,\ldots)$ & the infinite binary sequence\\\hline
$R^{(1-n)}$ & $r_{n-1}x^{1-n}+r_{n-2}x^{2-n}z+\cdots+r_0z^{1-n}$\\\hline
$\vv(F)$& the order of $F\in\M^\times$\\\hline\end{tabular}
\end{center}
\begin{center}
\begin{tabular}{|c|l|}\hline
Greek Symbol & Meaning \\\hline\hline
$\Delta_k=\Delta(f^{(k)};R^{(-1-k)})$ &  the discrepancy of  $f^{(k)}$ and $R^{(-1-k)}$ \\\hline
$\lambda_n$ & $|f^{(n-1)}|$\\\hline
$\varphi,\psi$ & forms in $\R^\times$\\\hline
$\varphi\circ F$ & form $\varphi$ acting on $F$\\\hline
$\rho$ & residue class of $Y$ in $\bbP$.\\\hline
\end{tabular}
\end{center}

\subsection{$\R$ and Grlex}
As in \cite{N15b}, we write $\succ$ for the graded-lexicographic order ({\em grlex}) on monomials of $\R^\times$, with $|x|=|z|=1$ and $x\succ z\succ1$. 
The leading term of a form $\varphi\in\R$ is written $\LT(\varphi)$. 
We define $\LL$ to be the set of {\em 'leading forms'}\,:
$$\LL=\{\varphi\in \R^\times: \varphi \ \mbox{is\ a\ form\ and}\ z\nmid\LT(\varphi)\}.$$

We also use $|\ |$ for the degree function on $\F[x]$, with $|0|=-\infty$.
Recall that the {\em homogenisation}  of $c\in\F[x]^\times $ is the form $c^\wedge\in\R$ given by $c(x,z)=z^{|c|}c(x/z)$ and the {\em dehomogenisation} of  $f\in\R[x,z]^\times\cap\LL$ is $f^\vee(x)=f(x,1)\in\F[x]$. Then
$|c^\wedge|=|c|$ and $|f^\vee|=|f|$.
 \subsection{Inverse Forms}
 {\em Throughout the paper, $F\in\M^\times$ denotes a  typical non-zero inverse form  of total degree $m=|F|\leq0$.}  
 
 We also order the  monomials of $\M^\times=\F[x^{-1},z^{-1}]^\times$ using grlex, now written $\prec$\,, but with $|x^{-1}|=|z^{-1}|=-1$ and $x^{-1}\prec z^{-1}\prec1$.

If $F$ is also a form i.e. an {\em inverse form},    we write $F=\sum_{j=m}^0F_{j,m-j}x^jz^{m-j}$;  {\em when $m$ is understood, we write $F_j$ for $F_{j,m-j}$}.  We will use a restriction of the exponential valuation $\vv$ for inverse forms: the {\em order} of $F$ is $\vv=\vv(F)=\max\{j: |F|\leq j\leq 0, F_j\neq0\}$. 
 The {\em augmentation of $F$ by $a\in\F$} is   $ax^{m-1}+Fz^{-1}$, an inverse form of total degree $m-1$.  For example,  the augmentation of $z^m$ by $a$ is $a\,x^{m-1}+z^{m-1}$.  
 A form $F$ defines  {\em inverse subforms}  by $\{F^{(j)}: m\leq j\leq \vv\}$ by $F^{(\vv)}=x^\vv$ and
  $$F^{(j)}=F_jx^j+F^{(j+1)}z^{-1}=F_jx^j+\cdots +x^\vv z^{j-\vv}\mbox{for}\ m\leq j\leq \vv-1.$$ 
We note that for any $j\leq 0$, $z^j(\circ F\,z^{-j})=F$. 
 
   {\em Throughout, $n\geq1$ and $(s_0,\ldots,s_{n-1})$ is a non-zero sequence (of elements of $\F$)}. 
    Thus $F=s_{n-1}x^{1-n}+\cdots+s_0z^{1-n}$  is the inverse form of $(s_0,\ldots,s_{n-1})$,  $s_{-\vv(F)}$ corresponds to the first non-zero term of the sequence, $a\,x^{m-1}+Fz^{-1}$ corresponds to the augmented sequence $(s_0,\ldots,s_{n-1},a)$ 
 and the inverse subforms of $F$ correspond to initial subsequences of 
 $(s_{-\vv(F)},\ldots,s_{n-1})$.

\subsection{The Module of Inverse Polynomials $\M$}
We recall the $\R$ module $\M=\F[x^{-1},z^{-1}]$ of { inverse polynomials} from \cite{Northcott}.  
\bd\label{mod} For  non-negative integers $p,q,u,v$ 
\begin{eqnarray}\label{module}
x^pz^q\circ x^{-u}z^{-v}=\left\{\begin{array}{ll}
x^{p-u}z^{q-v}&\mbox {if }p-u\leq 0, q-v\leq0\\
0&\mbox{otherwise.}
\end{array}
\right.
\end{eqnarray}
The $\R$ module structure of $\M$ is obtained by linearly extending 
(\ref{module})  to all of $\R$ and $\M$. 
\ed  By linearity, we can without loss of generality {\em assume that an inverse form $F$ satisfies $F_\vv=1$} i.e. $F=F_mx^m+\cdots +F_{\vv-1}x^{\vv-1}z^{m-\vv+1}+x^\vv z^{m-\vv}$. The module $\M$ is an $\R$ submodule of Macaulay's 'inverse system' $\F[[x^{-1},z^{-1}]]$ but we will not need this fact; see \cite{Northcott} for details.

The {\em annihilator ideal} of an inverse form $F$ is $$\I_F=\{\varphi\in\R: \varphi\circ F=0\}.$$

\bp (\cite[Proposition 3.7]{N15b}) \label{homog} The ideal $\I_F$ is homogeneous.
\ep
\bp\label{ann} (\cite[Lemma 3.1]{N15b}) For forms $\varphi\in\R$ and $F\in\M$ with $d=|\varphi|+|F|$ either (i) $d>0$ and  $\varphi\circ F=0$ or (ii) 
 $$\varphi\circ F=\sum _{j=d}^0[\varphi\cdot F]_j\,x^jz^{d-j}.$$
\ep
Thus we trivially have $x^n\in\I_F$ if $n>-|F|$.
\subsection{The Bijection}
Next we detail the bijection between  characteristic polynomials of a sequence and  the leading annihilating forms of its inverse form $F$ say. This uses the definition of a characteristic polynomial as in   \cite[Definition 2.7]{MR}, \cite{AD} (where sequences are indexed negatively) and \cite{Dai}. We use polynomial coefficients  in their natural order, not the reversed order of  'feedback coefficients'   and without using 'shift registers' of   the engineering literature. This definition enables us to exhibit a bijection between the set of characteristic polynomials of a sequence and the leading forms of the homogeneous ideal $\I_F$ of $\R$.

Let  $(s_0,\ldots,s_{n-1})$ be a sequence. Then
$c\in\F[x]^\times$  a {\em characteristic polynomial} of $(s_0,\ldots,s_{n-1})$ if $c$ is monic and either (i) $l=|c|\geq n$ or (ii) 
\begin{eqnarray}\label{char}
c_ls_{k+l}+\cdots+c_0s_k=0\ \mbox{for}\ 0\leq k\leq n-l-1.
\end{eqnarray}
We define the set
$$\chi(s_0,\ldots,s_{n-1})=\{c\in\F[x]^\times:c \mbox \ {is \ a \ characteristic \ polynomial\ of}\ (s_0,\ldots,s_{n-1})\},$$
which is not an ideal.

As $x^n$ is (vacuously) a characteristic polynomial of $(s_0,\ldots,s_{n-1})$, $\chi(s_0,\ldots,s_{n-1})$ is non-empty and
$$\lambda=\lambda(s_0,\ldots,s_{n-1})=\min \{|c|:\ c\in\chi(s_0,\ldots,s_{n-1})\}$$ 
is well-defined. Thus  {\em minimal polynomials} of $(s_0,\ldots,s_{n-1})$ i.e. characteristic polynomials of minimal degree $\lambda=\lambda(s_0,\ldots,s_{n-1})$
 are well-defined and is  known as the {\em linear complexity} of  $(s_0,\ldots,s_{n-1})$. We note that in \cite{Dai}, a characteristic polynomial of $(s_0,\ldots,s_{n-1})$ is written $c_n$ i.e. it is indexed by the {\em length} of the sequence, not its last term. 

Recall that $\LL$ is the set of monic, leading forms in $\R$ i.e. $\varphi$ such that  $z\nmid\LT(\varphi)$. If $F$ is an inverse form then  $x^{1-|F|}\in
\I_F^\times\cap\LL$. Thus  $\I_F^\times\cap\LL\neq\emptyset$ and we can consider forms in $\I_F^\times\cap \LL$ of minimal total degree. So we   define
$$\lambda_F=\min\{|f|:\ f\in\I_F^\times\cap\LL\}$$
and call $\lambda_F$ the {\em linear complexity of $F$}.
\bt \label{bij} Let  $(s_0,\ldots,s_{n-1})$ be a sequence with inverse form $F\in\M^\times$.  Then  $\wedge, \vee$ are mutual, degree-preserving bijections
$$\wedge: \chi(s_0,\ldots,s_{n-1})\leftrightarrows \I_F^\times\cap \LL: \vee$$
given by $\wedge(c)=c^\wedge$ and $\vee(f)=f^\vee$
so that for $c\in\chi(s_0,\ldots,s_{n-1})$, $|\wedge(c)|=|c|$ and for $f\in \I_F^\times\cap \LL$, $|\vee(f)|=|f|$. Therefore $\lambda(s_0,\ldots,s_{n-1})=\lambda(F)$.
\et
\bpr We have $|F|=1-n$ and $|c^\wedge|=|c|=l$. From Proposition \ref{ann}, $c^\wedge\circ F=0$ if and only if $[c^\wedge\cdot F]_j=0$ for $l+|F|\leq j\leq 0$.  Now  $c\in\chi(s_0,\ldots,s_{n-1})$ if and only if $c$  satisfies  (\ref{char}), and  substituting $k$ for $-j$, one sees that  Equation (\ref{char}) is equivalent to  $[c^\wedge\cdot F]_j=0$ for $l+|F|\leq j \leq 0$ i.e. equivalent to $c^\wedge\in\I_F\cap\LL$. This means that $c\in\chi(s_0,\ldots,s_{n-1})$ if and only if $c^\wedge\in\I_F^\times\cap \LL$, we have the required bijections  and hence  $\lambda(s_0,\ldots,s_{n-1})=\lambda(F)$.
\epr

\br \label{ctilde} Instead of  Equation (\ref{char}), \cite{Ma69} uses the equivalent $l$ and  'connection' polynomial 
$\gamma\in\F[x]^\times$ satisfying
\begin{eqnarray*}
\gamma_0s_j+\cdots+\gamma_ls_{j-l}=0\ \mbox{for}\ l\leq j\leq n-1
\end{eqnarray*}
where $\gamma_0=1$ and $\gamma_l$ may be zero (put $j=k+l$ and $\gamma(x)=\tilde{c}(x)=x^l\cdot c(x^{-1})$, the reciprocal of $c$, made monic). Then the  LFSR Synthesis Algorithm returns $l$ and $\gamma$.  Unfortunately, this formulation (using  reciprocal polynomials)  vitiates our algebraic approach. 
\er
The next section recalls how to construct a {\em generator} $f$ of $\I_F$ in $\LL$, so we automatically have a minimal annihilator of $F$ and hence a minimal polynomial $f^\vee$ of $(s_0,\ldots,s_{n-1})$. Minimal polynomials of negatively-indexed sequences and an algorithm to compute them and their corresponding numerators were studied in \cite{MR}.

\section{VOP's}\label{VOP}
\subsection{The Inductive Construction and Algorithm}
We recall the construction of a pair of generating forms  for $\I_F$ from \cite{N15b} and give an illustrative example of the algorithm at work. This construction is iterative, with a simple inductive  basis (Proposition \ref{001}) and an undemanding inductive step (Theorem \ref{constr}). 

 Let $F$ be an inverse form.  We say that an ordered pair of forms $(f,g)\in\R^2$ is a {\it viable ordered pair (VOP) for $\I_F$} if 
 
\noindent (i) $f,g$ are non-zero monic forms, $f\in\LL$ and $z|g$
 
\noindent  (ii)  $\I_F=\langle f,g\rangle$  (we call $f$   {\em a leading generator} and $g$  a {\em cogenerator} of $\I_F$)
 
\noindent  (iii) $|f|+|g|=2-|F|$.
 
\bp \label{001}(\cite[Proposition 3.8]{N15b}) If $F=x^m$ then  $\I_F=\langle x^{1-m},z\rangle$.
 \ep

The reader may check that $(f,g)=(x^{1-m},z)$ is a VOP for $\I_{\,x^m}$. 
If $f\in\I_F\cap\LL$, we call $f$ a {\em leading form}  for $\I_F$. 

Given a VOP $(f,g)$ for $\I_F$ and $G=ax^{m-1}+Fz^{-1}$ for some $a\in\F$, we need to know how to construct  a VOP  $(\varphi,\psi)$ for $\I_G$. This requires a notion of {\em 'discrepancy' } which shows how $a$ and $\I_F$ affect $\I_G$. It is our  analogue of  'discrepancy'  introduced in \cite{Ma69}; it is the obstruction to extending $f$ to a leading form  of $\I_G$.
\bd\label{discrepancy} {\rm If     $f\in\I_F^\times$ is a form and $G= ax^{m-1}+z^{-1}F$ then the {\em discrepancy} of $f$ and $G$ is
 $$\Delta(f;G)=[f\cdot G]_{(|f|+|G|,0)}\in\F\mbox{\ \ \ \  if }|f|+|G|\leq0$$ and $\Delta(f;G)=0$ otherwise.}
\ed

The inductive step is:
\bt \label{constr}(\cite[Proposition 4.6, Theorem 4.12]{N15b}) Let  $(f,g)$ be a VOP  for $\I_F$, $a\in\F$ and $G=ax^{|F|-1}+Fz^{-1}$. Suppose that $g\not\in\I_G$ and put  $\Delta'=\Delta(g;G)\in\F^\times$, $d=|g|-|f|$, $\Delta=\Delta(f;G)\in\F$.
If $\Delta=0$, set $(\varphi,\psi)=(f,zg)$ and $d=d+1$. 

But if $\Delta\neq0$, put  $q=\Delta/\Delta'$ and
\begin{eqnarray*}
(\varphi,\psi)=\left\{\begin{array}{ll}
(f-q\,x^{-d}g,z\,g)&\mbox {if }  d\leq0\\
(x^{+d}f-q\,g,zf)&\mbox{otherwise.}
\end{array}
\right.
\end{eqnarray*}
Then  $(\varphi,\psi)$ is a VOP for $G$
and manifestly $|\varphi|=|f|$ if $d\leq 0$ and $|\varphi|=|g|$ otherwise. In particular,
$|\varphi|=\max\{|g|,|f|\}$.\\

\noindent For the next iteration,
(i)  $e=|\psi|-|\varphi|=1$ if $\Delta=0$ and $e=1-|d|$ if $\Delta\neq0$ 
(ii) if $d>0$, $b\in\F$ and $\Delta(\varphi;b\,x^{|G|-1}+Gz^{-1})\neq0$,  we put $\Delta'=\Delta$, otherwise $\Delta'$ is unchanged. 
\et

\brs (i) A standard initialisation is $(f,g)=(1,0)$ giving $\I_{F^{(0)}}=\R$ if $F_0=0$, but  $|f|+|g|=2-m$ fails in this case. For this reason, we first find $\vv(F)$  and initialise with the construction with $(f,g)=(x^{1-\vv(F)},z)$. For example, $\I_{F^{(0)}}=\langle x, z\rangle$ if $F_0\neq0$.  

(ii) Given a form $F$, we iterate the construction to determine a VOP for $\I_{F^{(j)}}$ where $m\leq j\leq \vv$ and $F^{(m)}=F$.

(iii) An  iterative application of  the construction does not require  initialising with a particular VOP; we could initialise with any known VOP $(f,g)$ for $F^{(j)}$. 

(iv) If $\F=\GF(2)$ we can dispense with $\Delta'$.

(v) The reader may easily verify that we can write $(\varphi,\psi)=(f,g)M$ where $M$ is one of three possible $2\times 2$ matrices with coefficients in $\F[x]$, which depend on $\Delta$, $q$ and $d$. 
\ers
\bp (\cite[Proposition 3.5]{N15b}) If $(f,g)$ is a VOP for $F$ then $\lambda(F)=|f|$.
\ep 

We will often write $d_k=|g^{(k)}|-|f^{(k)}|$ and  $\Delta_k=\Delta(f^{(k)};F^{(-k-1)})$. When $\F=\GF(2)$, we will refer to $x^{d_k}$ (if $\Delta_k\neq0\ \mbox{and}\ \ d_k>0)$ and $x^{-d_k}$ (if $\Delta_k\neq0\ \mbox{and}\ d_k\leq0)$ as the {\em intermediate shifts} in the construction. 

We next recall \cite[Algorithm 4.22]{N15b}. This is an $\mathcal{O}(m^2)$ algorithm, \cite[Proposition 4.25]{N15b}, the proof of which is shorter and easier to understand than \cite{Gus}.
\begin{algorithm}\label{calPA}( $\F$ an arbitrary field)
\begin{tabbing}
\noindent {\tt Input}: \ \ \=  Inverse form $F\in\M^\times$.\\

\noindent {\tt Output}: \> VOP $(f,g)$ for $\I_F$.\\\\

(* Inductive basis: find $F^{(\vv)}$ and a VOP for $\I_F$ *)\\
$\lceil$ $j\la 1$; {\tt repeat}  $j\la j-1$  {\tt until} $(F_j\neq0)$; $\vv\la j$;\\ 
$ (f,g)\la (x^{1-\vv},z)$;\\\\

(* Inductive Step *)\\
 $\Delta'\la 1$;\ $G\la x^\vv$; $d \la  \vv$; \\\\
{\tt for} \= $j\la \vv-1$ {\tt downto }$|F|$ {\tt do}\\\\

   \>$\lceil$  $G\la F_jx^j+ Gz^{-1}$; $\Delta\la \Delta(f; G); q\la \Delta/\Delta'$;\\  \>{\tt if }$(\Delta\neq 0)$ {\tt then} \={\tt if} \=$(d\leq 0)$\ \={\tt then}\, $ f\la  f-q\,  x^{-d}\, g $; \\
  \> \> \>\> {\tt else}
   $\lceil$\= $t\la f$; $f \la  x^{+d}\, f-q\, g; g\la t$; \\
    \>\> \> \>\>$\Delta'\la  \Delta$; $d \la  -d;\rfloor$\\
   \> $g\la zg$; $d  \la  1+d$;$\rfloor$\\\\
{\tt return }$(f,g).\rfloor$
\end{tabbing} 
  \end{algorithm}
  
  We illustrate  Algorithm \ref{calPA} by revisiting  the sequence $(1,0,0,0,-1,1,0,0,1,-2,\ldots)\in\Q^{10}$  of  \cite[Example 4.8]{Fitz}, where it is regarded as the initial terms of a linear recurring sequence with a  minimal polynomial of degree at most 5.  We will  obtain its (unique) minimal polynomial $x^5+x-1$ without these assumptions.
  \be \label{Fitzex} { The sequence $(1,0,0,0,-1,1,0,0,1,-2)\in\Q^{10}$ defines the inverse form $F=-2x^{-9}+x^{-8}z^{-1}+x^{-5}z^{-4}-x^{-4}z^{-5}+z^{-9}\in\Q[x^{-1},z^{-1}]$. The subforms of $F$ are:   
\begin{center}
\begin{tabular}{|c|r|}\hline
$j$&$F^{(j)}$   \\\hline\hline
$0$&$1$   \\\hline
$-1$&$z^{-1}$    \\\hline
$-2$&$z^{-2}$  \\\hline
$-3$&$z^{-3}$      \\\hline
$-4$&$-x^{-4}+z^{-4}$   \\\hline
$-5$&$x^{-5}-x^{-4}z^{-1}+z^{-5}$   \\\hline
$-6$&$x^{-5}z^{-1}-x^{-4}z^{-2}+z^{-6}$  \\\hline
$-7$&$x^{-5}z^{-2}-x^{-4}z^{-3}+z^{-7}$    \\\hline
$-8$&$x^{-8}+x^{-5}z^{-3}-x^{-4}z^{-4}+z^{-8}$  \\\hline
$-9$&$-2x^{-9}+x^{-8}z^{-1}+x^{-5}z^{-4}-x^{-4}z^{-5}+z^{-9}$ \\\hline
\end{tabular}
\end{center}
 and we obtain the VOP $(f^{(9)},g^{(9)})$ for $\I_F$\,, as in the following summary:
\begin{center}
\begin{tabular}{|c|c|c|c|c|l|}\hline
$k$&$d_{k-1}$ & $\Delta'_{k-1}$ & $\Delta_{k-1}$ & $q_{k-1}$   &$(f^{(k)},g^{(k)})$    \\\hline\hline
$0$ &$-$ &$1$  &$-$  &$-$& $(x,z)$  \\\hline
$1$ &$0$  & $1$&$0$ &$0$ & $(x,z^2)$ \\\hline
$2$ &$1$   & $1$ &$0$&$0$  &$(x,z^3)$  \\\hline
$3$  &$2$  &$1$  &$0$&$0$  & $(x,z^4)$   \\\hline
$4$  &$3$ & $1$&$-1$ &$-1$  & $(x^4+z^4,xz)$ \\\hline
$5$  &$-2$&$-1$  &$1$ &$-1$& $(x^4+x^3z+z^4,xz^2)$  \\\hline
$6$ &$-1$  & $-1$ & $1$&$-1$ & $(x^4+x^3z+x^2z^2+z^4,xz^3)$ \\\hline
$7$   &$0$ &$-1$  &$1$& $-1$  & $(x^4+x^3z+x^2z^2+xz^3+z^4,xz^4)$\\\hline
$8$ &$1$ &$1$ &$1$ & $-1$  & $(x^5+x^4z+x^3z^2+x^2z^3+2xz^4,f^{(7)}z)$ \\\hline
$9$ &$0$ &$1$ &$1$ & $1 $  & $(x^5+xz^4-z^5,f^{(7)}z^2)$ \\\hline
\end{tabular}
\end{center}
so that $\I_F=\langle x^5+xz^4-z^5,x^4z^2+x^3z^3+x^2z^4+xz^5+z^6\rangle$.
Moreover, since $|g^{(9)}|=6>5=|f^{(9)}|$, $f^{(9)}=x^5+xz^4-z^5\in\Q[x,z]$ is the unique (monic) minimal leading form of $F$ and $\lambda(s)=|f^{(9)}|=5$. In particular, the unique minimal polynomial of the given finite sequence is $x^5+x-1\in\Q[x]$.
It  follows that the LFSR synthesis algorithm will return the reciprocal of $f^{(9)^\vee}$ made monic, namely $x^5-x^4-1$, as per Remark \ref{ctilde}.}
\eex

Likewise the methods of \cite{Be68},  \cite{Fitz} and \cite{LN}  do not apply to  the binary  sequence as we do not {\em a priori} know an upper bound $t$ for the linear complexity of the first $2t$ terms of the sequence.

 \subsection{Some Additional Properties}
If $m\leq0$ and $F=x^m$  then $\I_F=\langle x^{1-m},z\rangle$ by Proposition \ref{001} and $\gcd(x^{1-m},z)=1$. The next result shows that a constructed viable $(f,g)$ satisfies $\gcd(f,g)=1$.
\bp \label{gcd} If $(f,g)$ is a VOP for $\I_F$, $\gcd(f,g)=1$ and $\varphi,\psi$ are as constructed, then $\gcd(\varphi,\psi)=1$.
\ep
\bpr  If $\Delta=0$ and $h|\gcd(f,zg)$, then $h=z$ or $h|g$. But if $z|f$ then $z|\LT(f)$ which is impossible, so $h|\gcd(f,g)$. Suppose that $\Delta\neq0$ and $d\leq 0$.
Then $\gcd(\varphi,\psi)=\gcd(f,zg)=\gcd(f,g)=1$ as before. On the other hand, suppose that $d>0$. If $h|zf$ and $h|g$, then either (i) $h=z$ and $h|\varphi$ or (ii) $h|f$ and $h|(x^df-q\,g)$. But (i) is impossible since $z\nmid\LT(\varphi)$ so (ii) obtains and $h|\gcd(f,g)=1$.
\epr

If $m\leq0$ and $F=x^m$  then $(x^{1-m},z)$ is a VOP for $\I_F$, and $x^{1-m}$ is a leading form of minimal degree. But so is $x^{1-m}+\varphi \,z$ for any form with $|\varphi|=-m$. More generally,
\bc \label{theta} (Cf. \cite[Theorem 1]{Nied87}) Let  $(f,g)$ be a VOP for $\I_F$ and
$$\Theta=\{\theta\in\I_G\cap\LL  :\ |\theta|\ is\ minimal\}.$$ Then
\begin{eqnarray*}\Theta=\left\{\begin{array}{ll} 
\{ f\}				&\mbox{if}\ |g|>|f|\\
\{ f\}\cup\{ f+\psi\cdot g,  \psi\ \mbox{is\ a\ form\ and}\ |\psi|=|f|-|g|\}     & \mbox{otherwise.}
\end{array}\right.
 \end{eqnarray*}
\ec
\bpr Since $f$ and $g$ generate $\I_F$,  $f$ is the only monic leading annihilating form of minimal degree $|f|$ if  $|g|>|f|$, but if $|g|\leq |f|$ and $\psi\in\R^\times$ is a form with $|\psi|=|f|-|g|$ then $h=f+\psi\cdot g\in\I_F$ is a monic leading form since $\LT(h)=\LT(f)$ and $|h|=|f|$ is minimal.
\epr

In \cite[Example 4.24]{N15b}, we obtained $(f,g)=(x^4+x^3z+x^2z^2,x^3z+x^2z^2+xz^3+z^4)$ for the inverse form $F=x^{-6}+x^{-4}z^{-2}+x^{-3}z^{-3}+z^{-6}$. Here $f(0,1)=0$.  However, $h=f+g\in\I_F$ satisfies $h(0,1)=1$. (In fact, $h\in\I_{z^{-1}F}$.)

More generally, if we have an inverse form $F$ and begin iterating with $(x^{1-|F|},z)$,  then the construction provides a VOP $(f,g)$ with $\gcd(f,g)=1$. Hence if $f(a,b)=0$ for some $a,b\in \F$ then $g(a,b)\neq0$ and if $|g|\leq |f|$, then $h=f+x^{|f|-|g|}g$ is a leading form in $\I_F$ such that $h(a,b)\neq0$. On the other hand, if $|g|> |f|$, then $h=x^{|g|-|f|}f+g\in\I_F$ is a leading form  such that $h(a,b)\neq0$, but of increased degree $|g|$. 

We recall the following result from \cite{N15b} for later reference:
\bt \label{dehomog} Let $F$ be the inverse form for $(s_0,\ldots,s_{n-1})$, $(f,g)$ be the VOP  for $\I_F$ as constructed above and let $(\mu,\nu)$ be the minimal polynomial pair for $(s_0,\ldots,s_{n-1})$ as constructed in \cite[Algorithm 6.14]{N15b}. There is a 1-1 correspondence $(f,g)\mapsto (f^\vee,g^\vee)$ with  and  
$(\mu,\nu)\mapsto (\mu^\wedge,\nu^\wedge z^{n+1-|\mu|-|\nu|})$ for $\nu\neq0$.
\et

\brs (i) We trivially have $|\mu^\wedge|+|\nu^\wedge z^{n+1-|\mu|-|\nu|}|=n+1$, which  agrees with VOP Property (iii) since $m=1-n$. 

(ii) $|g|>|f|\Leftrightarrow n+1-|\mu|>|\mu|$ and $\max\{|g|,|f|\}=\max\{n+1-|\mu|,|\mu|\}$, cf. \cite{Ma69}. 

(iii) Theorem \ref{dehomog} means that we can also compute a pair of generators for $\I_ {(s_0,\ldots,s_{n-1})}$ using \cite[Algorithm 3.14]{MR}.
\ers
\subsection{Linear Complexity Profiles}
An inverse form $F$ has subforms $F^{(j)}$ for $m=|F|\leq j\leq \vv=\vv(F)$, with $F^{(\vv)}=x^\vv$ and $F^{(m)}=F$. We write 
 $\lambda(F^{(j)})$ for the linear complexity of  the subform $F^{(j)}$.

 We call the  sequence $(\lambda(F^{(\vv)}),\ldots, \lambda(F^{(m)}))$ of integers the {\em linear complexity profile} of $F$. We will say that $F$ has a  {\em a perfect linear complexity profile (PLCP)} if $\vv=0$ and $\lambda(F^{(-k)})=\lfloor (k+2)/2\rfloor$ for $0\leq k\leq-m$. From Theorem \ref{ann}, this agrees with the usual notion of the linear complexity profile of a sequence.
Next  we  relate the notion of PLCP  to the intermediate shifts occurring in Theorem \ref{constr}. This is an analogue of \cite[Theorem 2]{Nied87}; however we do not  require an irrationality hypothesis.
 
\bp \label{PLCP} Let $F$ be an inverse form with $F_0=1$. For $1\leq k<-m$, let $(f^{(k)},g^{(k)})$ be a VOP for $\I_{F^{(-k)}}$ and $F^{(-1-k)}$ be the $(-1-k)^{th}$ subform  of $F$. Put $\Delta_k=\Delta(f^{(k)};F^{(-1-k)})$. The following are equivalent:

(i)  $F$ has a PLCP

(ii) the intermediate shift is $x$ if and only if $k$ is odd
\ep
\bpr  The quantity $\lfloor (k+1)/2\rfloor$ is 1 for $k=1$ and increases by 1 if and only if $k$ is odd. Since $F_0=1$, $(f^{(0)},g^{(0)})=(x,z)$ i.e. $|f^{(0)}|=1$ and $d_0=0$. Thus either $(f^{(1)},g^{(1)})=(x,z^2)$ or $(f^{(1)},g^{(1)})=(x+z,z^2)$. Next, the degree $|f^{(k)}|$ increases by 1 if and only if $k$ is odd is equivalent to 
$f^{(k+1)}=xf^{(k)}+q_k g^{(k)}$ if $k$ is odd and $f^{(k+1)}=f^{(k)}+q_k g^{(k)}$ if $k$ is even.  
\epr
\br In Proposition \ref{PLCP},  $\Delta_k\in\GF(2)$ is arbitrary when $k$ is even. The expected linear complexity of a random binary sequence of length $n$ is $n/2+c_n$ where 
$0\leq c_n\leq 5/18$, \cite[Ch. 4]{R86}. Thus a binary sequence with $s_0=1$ and $s_i$ chosen so that $\Delta_k=1$ when $k$ is odd and randomly when $k$ is even will have  a PLCP and be a good approximation to  a random binary sequence.
\er
\section{Rueppel's Conjecture}\label{GF2}
{\em For the remainder of the paper, $\F=\GF(2)$.} Rueppel's conjecture concerns the  binary sequence   $r=(r_0,r_1,\ldots)=(1,1,0,1,0^3,1,0^7,1,0^{15},1,\ldots)$ 
where $r_i=1\in\F$ if $i=2^k-1$ for some $k\geq0$ and $0$ otherwise.

\bd For $n\geq1$, the  inverse form of $(r_0,\ldots,r_{n-1})$  is 
$$R^{(1-n)}=R^{(1-n)}(x^{-1},z^{-1})=\sum_{j=1-n}^0 r_{-j}x^jz^{1-n-j}\in\F[x^{-1},z^{-1}]$$
and we write $\I_{n-1}$ for $\I_{R^{(1-n)}}$.
\ed
Note that $R^{(1-n)}$ and $\I_{n-1}$ are indexed using the  last index of $(r_0,\ldots,r_{n-1})$ rather than the length of the sequence.
We have $|R^{(1-n)}|=1-n$,  $R^{(1-2^k)}=\sum_{j=k}^0 x^{1-2^j}z^{2^j-2^k}$ for $k\geq 0$, and if  $k\geq1$ then $R^{(1-2^k)}=x^{1-2^k}+R^{(1-2^{k-1})}z^{-2^{k-1}}$. In addition, if $2^k-1<n<2^{k+1}-1$ then $R^{(-n)}=R^{(1-2^k)}z^{2^k-1-n}\in z^{-1}\M$.
\be\label{rR} 
The inverse forms $R^{(j)}$ for $-9\leq j\leq 0$ are as follows:
\begin{center}
\begin{tabular}{|r|rl|}\hline
$j$  &$R^{(j)}$ & \\\hline\hline
$0$ &$1$                                 & $=R^{(0)}$   \\\hline
$-1$ &$x^{-1}+z^{-1}$               &$=R^{(-1)} $  \\\hline\hline
$-2$ &$x^{-1}z^{-1}+z^{-2}$      &$=R^{(-1)}z^{-1}$      \\\hline
$-3$ &$x^{-3}+{x^{-1}z^{-2}}+z^{-3}$    &$=R^{(-3)}$   \\\hline\hline
$-4$ &$x^{-3}z^{-1}+x^{-1}z^{-3}+z^{-4}$ &$=R^{(-3)}z^{-1}$    \\\hline
$-5$ &$x^{-3}z^{-2}+x^{-1}z^{-4}+z^{-5}$  &$=R^{(-3)}z^{-2}$   \\\hline\hline
$-6$ &$x^{-3}z^{-3}+x^{-1}z^{-5}+z^{-6}$  &$=R^{(-3)}z^{-3}$   \\\hline
$-7$&$x^{-7}+x^{-3}z^{-4}+x^{-1}z^{-6}+z^{-7}$  &$=R^{(-7)}$ \\\hline
$-8$&$x^{-7}z^{-1}+x^{-3}z^{-5}+x^{-1}z^{-7}+z^{-8}$  &$=R^{(-7)}z^{(-1)}$ \\\hline
$-9$&$x^{-7}z^{-2}+x^{-3}z^{-6}+x^{-1}z^{-8}+z^{-9}$  &$=R^{(-7)}z^{(-2)}$. \\\hline
\end{tabular}
\end{center}
\eex
 Recall that $\Delta_{k-1}=\Delta(f^{(k-1)};R^{(-k)})=[f^{(k-1)}\cdot R^{(-k)}]_{(|f^{(k-1)}|-k,0)}$  and  $q_{k-1}=\Delta_{k-1}$ for $0\leq k-1\leq n-1$.

\be \label{first8} (Ex. \ref{rR} cont.)
 Firstly, since $R^{(0)}=1$, $\I_0=\langle f^{(0)},g^{(0)}\rangle=\langle x,z\rangle$ from Proposition \ref{001}.  However, we can and will take $(f^{(0)},g^{(0)})=(x+z,z)$ so that $f^{(0)}(0,1)=1$. The key ingredients for the construction are  the degree increment $d_{k}=|g^{(k)}|-|f^{(k)}|$ and the discrepancy  $\Delta_k$ -- so that we know how to update $(f^{(k)},g^{(k)})$. We obtain the following table:
\begin{center}
\begin{tabular}{|c|c|c|l|l|}\hline
$k$&$d_{k-1}$  & $\Delta_{k-1}$     &$f^{(k)}$  &$g^{(k)}$ \\\hline\hline
$0$	&$-$ &$-$  & $f^{(0)}=x+\ul{z}$&$z$ \\\hline
$1$	&$0$  & $0$ & $f^{(0)}$&$z^2$   \\\hline\hline
$2$	&$1$    & $1$    &$xf^{(1)}+g^{(1)}=x^2+\ul{xz}+z^2$&$f^{(0)}z$  \\\hline
$3$	&$0$    &$0$   & $f^{(2)}$&$f^{(0)}z^2$  \\\hline\hline
$4$	&$1$   & $1$  & $xf^{(3)}+g^{(3)}=x^3+x^2z+\ul{z^3}$&$f^{(2)}z$\\\hline
$5$	&$0$  &$0$   & $f^{(4)}$&$f^{(2)}z^2$ \\\hline\hline
$6$	&$1$    & $1$  & $xf^{(5)}+g^{(5)}=x^4+\ul{x^3z}+x^2z^2+z^4$&$f^{(4)}z$ \\\hline
$7$	&$0$    &$0$ &  $f^{(6)}$&$f^{(4)}z^2$ \\\hline\hline
$8$	&$1$    & $1$  & $xf^{(7)}+g^{(7)}=x^5+x^4z+\ul{x^2z^3}+xz^4+z^5$&$f^{(6)}z$ \\\hline
$9$	&$0$    &$0$ &  $f^{(8)}$&$f^{(6)}z^2$. \\\hline
\end{tabular}
\end{center}
Then for $0\leq k\leq 9$, 
$f^{(k)}=\lfloor (k+2)/2\rfloor=\lambda_{k+1}$ (recall that $f^{(k)}\circ (r_0,\ldots,r_k)=0$ where there are $k+1$ terms). As per Corollary \ref{theta},  $f^{(k)}$ is the unique leading annihilating form of minimal total degree if $k$ is odd and $f^{(k)}+g^{(k)}$ is the only other leading annihilating form of minimal total degree when $k$ is even. 
The significance of the underlined terms will become clear in the proof of Theorem \ref{main}. \eex 

We now come to our main result.
\bt \label{main}  Let $(f^{(0)},g^{(0)})=(x+z,z)$ and for $k\geq0$, let $(f^{(k+1)},g^{(k+1)})$ be as constructed in Theorem \ref{constr}. Then
 
(A) If $k$ is even then $\Delta_k=0$, otherwise $\Delta_k=1$ and $d_k=1$

(B) \begin{eqnarray*}
(f^{(k+1)},g^{(k+1)})=\left\{\begin{array}{ll}
(f^{(k)},zg^{(k)})&\mbox {if } \ k\ \mbox{is even}\\
(xf^{(k)}+g^{(k)},zf^{(k)})&\mbox{otherwise}
\end{array}
\right.
\end{eqnarray*}

(C) $|f^{(k+1)}|=\lfloor (k+3)/2\rfloor$

(D) $f^{(k+1)}(0,1)=g^{(k+1)}(0,1)=1$.
\et
\bpr   We apply Theorem \ref{constr} and Proposition \ref{PLCP} inductively, beginning with $(f^{(0)},g^{(0)})=(x+z,z)$ where
$\Delta_0=0$, $|f^{(0)}|=1=\lfloor 2/2\rfloor$ and $f^{(0)}(0,1)=1=g^{(0)}(0,1)$.

Suppose inductively that  the result is true for $k$. 

(A) For $f^{(k+1)}$, we have to determine $\Delta_k=[f^{(k)}\cdot R^{(-1-k)}]_{(|f^{(k)}|-1-k,0)}$. 
Let  $P=2^p-1\leq k+1<2^{p+1}-1$ for some $p\geq1$, put $e=|f^{(k)}|$ and $l=e-1-k$. Then
$$S=R^{(-1-k)}=\left(\sum_{j=p}^0 x^{1-2^j}z^{2^j-2^p}\right)z^{P-1-k}\ \mbox{and}\  
\Delta_k=[f^{(k)}\cdot S]_{(l,0)}.$$ 
We consider three cases: (i) $k$ even, $P=k+1$ (ii) $k$ even, $P<k+1<2^{p+1}-1$ and (iii) $k$ odd and $P\leq k<2^{p+1}-1$.  

(i) We have $k+1=P$ and have to show that $\Delta_k=0$. Since $k-1$ is odd, the inductive hypothesis gives $\Delta_{k-1}=1$, $d_{k-1}=1$ and $(f^{(k)},g^{(k)})=(xf^{(k-1)}+g^{(k-1)}, zf^{(k-1)})$. Also $|f^{(k)}|=|f^{(k-1)}|+1= k/2+1=2^{p-1}$, so $l=|f^{(k)}|-k-1=2^{p-1}+1-2^p=1-2^{p-1}$. By part (A) $f^{(k)}(0,1)=1$, so we can write $f^{(k)}\cdot S$ as
\begin{eqnarray*}
(x^e+\alpha+z^e)\cdot (x^{1-2^p}+x^{1-2^{p-1}}z^{-2^{p-1}}+\beta).
\end{eqnarray*}
 Then $x^e\cdot x^{1-2^p}=x^l$, and one checks that $(\alpha+z^e)\cdot (x^{1-2^{p-1}}z^{-2^{p-1}}+\beta)=x^l$ plus terms   in $z^{-1}\M$, so 
 $\Delta_k=[f^{(k)}\cdot S]_{(l,0)}=0$ and $f^{(k+1)}=f^{(k)}$.

(ii) Here $k$ is even, $P-k-1<0$ and $S\in z^{-1}\M$. As in case (i), $(f^{(k)},g^{(k)})=(xf^{(k-1)}+g^{(k-1)},zf^{(k-1)})$ and $e=|f^{(k)}|=|f^{(k-1)}|=2^{p-1}$. Then 
$$f^{(k)}\cdot S=(x^e+(f^{(k)}+x^e))\cdot R^{(-P)}z^{P-k-1}$$
 $x^e\cdot R^{(-P)}z^{P-k-1}$, $(f^{(k)}+x^e)\cdot x^{-P}$,
 $(f^{(k)}+x^e)\cdot (R^{(-P)}+x^{-P})z^{P-k-1}\in z^{-1}\M$, so $\Delta_k=0$ and again $f^{(k+1)}=f^{(k)}$. 

(iii) Next we consider $k$ odd. For $k=1$, $\Delta_1=[(x+\ul{z})\cdot (x^{-1}z^{-1}+z^{-2})]_{(1-2,0)}=1$; the term $\ul{z}$ of $f^{(1)}$ has triggered $\Delta_1=1$. The reader may easily verify that $\ul{xz}=\alpha z$ triggers $\Delta_3=1$ and that $\ul{z}^3=\beta z$ triggers $\Delta_5=1$. Now let $k$ be odd and $p\geq3$,  
$P=2^p-1\leq k<2^{p+1}-1$. Define maps $\alpha,\beta:\R\ra\R$ by $(\alpha\ f)(x,z)=x\cdot f(x,z)$ and 
$(\beta\ f)(x,z)=z^2\cdot f(x,z)$.
Since $k-1$ is even, the inductive hypothesis gives 
$$f^{(k)}=f^{(k-1)}=xf^{(k-2)}+ g^{(k-2)}=\alpha\ f^{(k-2)}+ \beta\ f^{(k-4)},$$ 
$e=|f^{(k)}|=|f^{(k-1)}|=(k+1)/2$ and $l=e+|S|=(k+1)/2-1-k=-(k+1)/2$. 

Put   $t_k= x^{u_k}z^{v_k}$ where $u_k=P-\frac{k+1}{2}$, $v_k=k+1-P$. Then $|t_k|=(k+1)/2=|f^{(k)}|$ and $t_k=\alpha^{u_k}\beta^{(v_k-1)/2}z$. Since $z$ is a term of $f^{(1)}$, $t_k$ is a term of  $f^{(2u_k+2v_k-1)}=f^{(k)}$.

Let $L=\LT(S)$ where $S=R^{(-P)}z^{P-k-1}\in z^{-1}\M$. Then
 \begin{eqnarray*} f^{(k)}\cdot S=f^{(k)}\cdot (L+S)+f^{(k)}\cdot L=f^{(k)}\cdot (L+S)+t_k\cdot L+(f^{(k)}+t_k)\cdot  L
 \end{eqnarray*}
and $t_k\cdot L=x^{(k-1)/2}z\cdot x^{-k}z^{-1}=x^l$. 
It is straightforward that $f^{(k)}\cdot (L+S)\in z^{-1}\M$ and $(f^{(k)}+t_k)\cdot  L\in z^{-1}\M$. Thus $\Delta_k=[f^{(k)}\cdot S]_{(l,0)}=[t_k\cdot L]_{(l,0)}=1$,  the term $t_k$ of $f^{(k)}$ triggers $\Delta_k=1$ and   $(f^{(k+1)},g^{(k+1)}) =(xf^{(k)}+g^{(k)},zf^{(k)})$. 

(B) This is  a simple consequence of Part (A).

(C) Suppose that  $|f^{(k)}|=\lfloor (k+2)/2\rfloor$. From Part (C), if $k$ is even then $|f^{(k+1)}|=|f^{(k)}|=(k+2)/2=\lfloor (k+3)/2\rfloor$ and if $k$ is odd then $|f^{(k+1)}|=|f^{(k)}|+1=\lfloor (k+2)/2\rfloor+1=(k+1)/2+1=(k+3)/2=\lfloor (k+3)/2\rfloor$. 

(D) We know that $f^{(k)}(0,1)=g^{(k)}(0,1)=1$ for $k=0,1$, so suppose that the result is true for $k$. If $k$ is even, $f^{(k+1)}(0,1)=f^{(k)}(0,1)=1$ and $g^{(k+1)}(0,1)=1$ whereas if $k$ is odd, $f^{(k+1)}(0,1)=g^{(k)}(0,1)=1$  and $g^{(k+1)}(0,1)=f^{(k)}(0,1)=1$.
\epr

\brs (i) For a field with more than two elements, cancellation may fail in the proof of Part (B) when $k$ is even. 
(ii) In Example \ref{first8}, the underlined terms  trigger a discrepancy of 1:  for odd $k$ and  $2^p-1\leq k< 2^{p+1}-1$,  $u_{k+1}=u_k-1$,  $v_{k+2}=v_k+2$, so that the $t_k$ in the proof of Theorem \ref{main} take the values $x^pz^1,x^{p-1}z^3,x^{p-2}z^5, \ldots, x^0z^{2^p-1}$. 
\ers

\bc Define $2\times 2$ matrices with entries in $\R$ by
\begin{eqnarray*}E=\left[\begin{array}{ll} 1 &0\\
                                                                            0 & z
                                              \end{array}\right]\ \mbox{and}\ 
U=\left[\begin{array}{ll} x &z\\
                                                                            1 & 0
                                              \end{array}\right].
 \end{eqnarray*}
  Then  (i) for $n\geq0$, $(f^{(n+1)},g^{(n+1)})=(f^{(n)},g^{(n)})M_n$ where $M_n=E$ if $n$ is even and $M_n=U$ if $n$ is odd
 
 (ii) for $i\geq 1$ and $P=UE$  
 $$(f^{(2i)},g^{(2i)})=(x+z,z^2)P^{i-1}U\ \mbox{and}\ 
 (f^{(2i+1)},g^{(2i+1)})= (x+z,z^2)P^i.$$
 
 (iii) if $\phi^{(i)}(x)=f^{(i)}(x,1)$ then $\phi^{(2i)}=(x+z,z^2)U^i$ for $i\geq1$.
 \ec
 \bpr Part (i) is a restatement of  Theorem \ref{main}(B). Part (ii) is a simple induction.
 \epr
 The next result is an immediate consequence of Proposition \ref{PLCP}  and Theorem \ref{main}.
\bc The sequence $(r_0,r_1,\ldots)$ has a PLCP.
\ec
\bpr For $n\geq1$,  $\lambda(r_0,r_1,\ldots,r_{n-1})=|f^{(n-1)}|=\lfloor (n+1)/2\rfloor$.
\epr
\bc Let $n\geq 0$, $( f^{(n)},g^{(n)})$ be as constructed for $\I_n$ and  $\Theta_n$ be the set $\{\theta\in\I\ \ is\ a\ leading\ form, |\theta|\ is\ minimal\}$. Then
\begin{eqnarray*}\Theta_n=\left\{\begin{array}{ll} 
\{f^{(n)}\}				&\ \mbox{if}\ $n$\ \mbox{is odd}\\
\{f^{(n)},f^{(n)}+g^{(n)}\}     &\ \mbox{otherwise.}
\end{array}\right.
 \end{eqnarray*}
\ec
\bpr This is Corollary \ref{theta} specialised to $R^{(-n)}$.
\epr

We next give a closed-form expression for $f^{(2k-1)}$ dehomogenised; this also motivates the use of the roots $Y^2+xY+1$ in a quadratic extension of $\F(x)$.

\bl \label{hk}  If $h^{(k)}\in\F[x]$, $h^{(1)}=x+1$, $h^{(2)}=x^2+x+1$ and  $h^{(k)}+x\,h^{(k-1)}+h^{(k-2)}=0$ for $k\geq3$ then  $$x\, h^{(k)}= (1+\rho)\rho^k+(1+\rho^{-1})\rho^{-k}$$
where $\rho=\ol{Y}\in \F(x)/(Y^2+xY+1)=\K$. 
\el
\bpr The polynomial $h^{(k)}$ has characteristic polynomial $Y^2+xY+1\in\F(x)[Y]$, which is irreducible -- one easily shows that if $Y^2+xY+1=(Y+u)(Y+v)$ for some $u,v\in\F(x)$ then $x=0$. So let $\rho=\ol{Y}\in\K$. Solving 
$h^{(k)}=A\rho^k+B\rho^{-k}$ for $A,B\in\F(x)$ subject to $h^{(1)}=x+1$, $h^{(2)}=x^2+x+1$ gives the required coefficients. 
\epr
 
 \bt \label{Hah} For $k\geq 1$ we have
$$x\,f^{(2k-2)}(x,1)=x\,f^{(2k-1)}(x,1)=(1+\rho)\rho^k+(1+\rho^{-1})\rho^{-k}=x\mu_{2k-1}(x)$$
where $\mu_{2k-1}$ was defined in Theorem \ref{dehomog}.
\et
\bpr  Theorem \ref{main} implies that we may take $h^{(k)}=f^{(2k-2)}(x,1)$ in Lemma \ref{hk}, $f^{(2k-2)}=f^{(2k-1)}$ and $|g^{(2k-1)}|=|f^{(2k-1)}|+1$, so 
$f^{(2k-1)}$ is unique. The last equality follows from Theorem \ref{dehomog}.
 \epr
 \bc (Cf. \cite[Lemma 5]{Dai}) Let $\eta=(1+\rho)\rho^k+(1+\rho^{-1})\rho^{-k}\in\K$ as in Lemma \ref{hk}. Then $\eta\in\F[x]$, $x|\eta$ and $|\eta|=k+1$.
 \ec
\bc \label{lfsr} The LFSR synthesis algorithm  applied to $(r_0,\ldots,r_{2k-1})$ returns $k$ and the reciprocal  of $f^{(2k-1)}(x,1)$.  
\ec
\bpr  From Theorem \ref{main}, $|f^{(2k-1)}(x,1)|=k$ and $f^{(2k-1)}(0,1)=1$.\epr

Theorem \ref{main} also shows that Algorithm \ref{calPA} applied to $R^{(1-n)}$ simplifies considerably : there is no need to multiply to find the discrepancy and little besides a parity test.
\begin{algorithm}\label{Ralg}  VOP algorithm simplified for $R^{(1-n)}$
\begin{tabbing}
\noindent {\tt Input}: \ \ \=  Integer $n\geq1$.\\

\noindent {\tt Output}: \>  Viable ordered pair $(f,g)$ for $\I_{n-1}$.\\

$\lceil$\=  $(f,g)\la (x+z,z)$;\\\\

\>{\tt for} \= $j\la 0$ {\tt downto }$1-n$ {\tt do}\\
          \> \>$\lceil$  \mbox{if}\ $j$\  \mbox{is odd}\ \={\tt then} $\lceil$ $t\la f$; $f \la  xf+g$; $g\la t$; $\rfloor$\\
          \>\>  $g\la z\,g$; $\rfloor$; \\\\
\>{\tt return }$(f,g).\rfloor$
\end{tabbing} 
  \end{algorithm}
  
  For additional properties of $\I_{n-1}$, e.g. its codimension and how to compute its (unique) reduced grlex Groebner Basis, we refer the reader to \cite{N15b}, \cite{N19}.

\section{The  Minimal Polynomials  of \cite{Dai}}\label{EADai}
Let $k\geq1$ and suppose given a sequence $(s_0,\ldots,s_{2k-1})$. In \cite{Dai}, the author constructs  $c_k\in\F[x]$ for  $(s_0,\ldots,s_{2k-1})$ using the EA: put  $s'_{-1}=x^k$, $s'_0=s_0x^{k-1}+\cdots+s_{k-1}$ and let $q_k, s'_k,c_k\in\F[x]$ be defined by $c_{-1}=0,c_0=1$,
$$ s'_k=q_ks'_{k-1}+s'_{k-2\ }\mbox{and}\ |s'_k|<|s'_{k-1}|,\ c_k=q_kc_{k-1}+c_{k-2}\,,
$$
terminating at the first $k$ such that $s'_k=0$. This is applied to the first $2k-1$ terms of $(1,1,0,1,0^3,\ldots)$ {\em with remainders} $r'_k$ say. The author adopts a different approach to finding the $c_k$, which we now summarise. The matrix
\begin{eqnarray*}V=\left[\begin{array}{ll} 1 &0\\
                                                                            0 & x
                                              \end{array}\right]
 \end{eqnarray*}
 has characteristic polynomial $Y^2+xY+1$ with roots $\rho^{\pm 1}$ as in Lemma \ref{hk} and characteristic vectors $[1\ \rho^{\pm 1}]^\mathrm{T}$; diagonalising $V^k$ yields   
 $q_1(x)=x+1$ and $q_k(x)=x$ for $k\geq2$, \cite[Lemmas 2, 3:3]{Dai}. This  implies that  $r'_k=x\,r'_{k-1}+r'_{k-2}$ for $k\geq2$. Further diagonalising  in \cite[Lemma 4]{Dai} together with \cite[Lemma 1]{Dai} yields a formula for $c_k$. 
 
 However, the key facts that $q_1(x)=x+1$ and $q_k(x)=x$ for $k\geq2$ give  $c_1(x)=x+1$, $c_2(x)=x^2+x+1$ and $c_k=x\,c_{k-1}+c_{k-2}$ for $k\geq2$. We can therefore apply Lemma \ref{hk}  to $c_k$.  This  together with Theorem \ref{main}   gives
 \bc \label{omnibus} For $k\geq1$
 
  (i) $c_k(x)=f^{(2k-1)}(x,1)=f^{(2k-2)}(x,1)=x^{-1}((1+\rho)\rho^k+(1+\rho^{-1})\rho^{-k})$
 
 (ii) $|c_k|=|f^{(2k-1)}(x,1)|=k$
 
 (iii)  $c_k$ is the unique minimal polynomial of $(r_0,\ldots,r_{2k-1})$

 (iv) $f^{(2k-1)}(x,z)=z^kc_k(x/z)$.
 \ec
As in \cite{Dai}, Lemma 1, {\em op. cit.}  implies that if $k\geq0$ and $l=2^k$, then $c_l(x)= x^l+\sum_{j=0}^{\log_2 l}x^{l-2^j}$ and since $c_l(x)=f^{(2l-1)}(x,1)$,
  $$f^{(2l-1)}(x,z)=c_l^{\ \wedge}(x,z)=x^l+\sum_{j=0}^{\log_2 l}x^{l-2^j}z^{2^j}.$$

\end{document}